\documentclass[12pt,a4paper]{article}
\usepackage{amsmath}
\usepackage{amsthm}
\usepackage{amssymb}
\usepackage[left=1cm,right=1cm,top=2cm,bottom=2cm]{geometry}

\def\be{\begin{eqnarray}}
\def\ee{\end{eqnarray}}

\begin{document}
\hfill ITEP-TH-50/12

\bigskip

\centerline{\Large{The first-order deviation of superpolynomial}}
\centerline{\Large{in an arbitrary representation from the special polynomial}}

\bigskip

\centerline{Anton Morozov\footnote[1]{anton.morozov@itep.ru}}

\bigskip

\centerline{{\it Moscow State University and ITEP, Moscow, Russia}}

\bigskip

\centerline {ABSTRACT}

\bigskip

{\footnotesize
Like all other knot polynomials, the superpolynomials \cite{DGR}
should be defined in arbitrary representation R of
the gauge group in (refined) Chern-Simons theory \cite{CS,CSref}.
However, not a single example is yet known of
a superpolynomial beyond symmetric or antisymmetric
representations.
Following \cite{MMS}, we consider the expansion of
the superpolynomial around the special polynomial
in powers of $q-1$ and $t-1$
and suggest a simple formula for the first-order
deviation, which is presumably valid for arbitrary
representation.
This formula can serve as a crucial lacking test
of various formulas for non-trivial superpolynomials,
which will appear in the literature in the near future.
}

\bigskip

As it was shown in the article \cite{M}, some superpolynomials in the case of symmetrical and antisymmetrical  representations possess simple factorization properties in the case of $q=1$ and $t=1$ respectively,
which extend the corresponding property of the ordinary special polynomial \cite{zhu,DMMSS}.
Now this property was also checked in \cite{inds} for all twisted knots, but there are arguments, that
it does not hold for some more complicated knots \cite{GS}.

There can be three directions to continue this research:

-- to look at the other knots,

-- to see what happens if $q$ or $t$ deviate from unity: $q=1+\hbar+\ldots$ or $t=1+\bar\hbar+\ldots$

-- and to look at arbitrary representations.

The third direction is most interesting, but the problem is that we do not have any examples
of superpolynomials even in the case of representation $[2,1]$ and as a result we can not really check
our conjectures. Thus we choose a way in between: try to imagine, what the answer could be
for the infinitesimal deviation from special polynomials, but for arbitrary representation
and, perhaps, for generic knots.

Let us parameterize  the small deviations of $q$ and $t$ from unity as follows:
\be
q = e^{\hbar}; \,\,\,\, t = e^{\overline{\hbar}}
\ee
In this parametrization our superpolynomial can be written as $P_R (A,\hbar,\overline{\hbar})$.
The special polynomial arises at $\hbar = 0$ or $\overline{\hbar} = 0$.
and it satisfies \cite{DMMSS}
\be
P_R (A, 0, 0)=H_R(A,0)=\left(H_{{_\square}}(A,0)\right)^{|R|}
\ee
where ${\square}$ denotes the fundamental representation,
and $|R|$ is the number of boxes in the Young Diagram of representation $R$.

In the next approximation we have:
\be
P_R (A,\hbar,\overline{\hbar}) = \sigma_{{_\square}}^{|R|}(A) +\hbar \eta_R(A) + \overline{\hbar} \overline{\eta}_R(A) + \ldots
\ee
where $\sigma_{{_\square}}=H_{{_\square}} (A,0)$, and
$\eta_R =  \left.\frac{\partial\left(P_R (A,\hbar,0)\right)}{\partial \hbar}\right|_{\hbar = 0}$,
$\overline{\eta}_R = \left.\frac{\partial\left(P_R (A,0,\overline{\hbar})\right)}{\partial \overline{\hbar}}\right|_{\bar{\hbar = 0}}$.

\bigskip

Now let us see what can be said about the functions $\eta$ and $\bar \eta$.
For symmetric representations we use the factrorization property \cite{M} of
the special {\it super}polynomial (assuming that it is true for our knot):
\be
P_{[r]}(A,0,\bar\hbar)=\Big(P_{{_\square}}(A,0,\bar\hbar)\Big) ^r
\ee
This relation is conjecturally true for all $\bar\hbar$, but we need it only in the first order,
when it implies:
\be
\overline{\eta}_{[r]}(A)= r \sigma_{{_\square}}^{r-1} (A) \overline{\eta}_{{_\square}}(A)
\ee
Similarly \cite{M}, for antisymmetric representation
\be
P_{[1^r]}(A,\hbar,0)=\Big(P_{{_\square}}(A,\hbar,0)\Big)^r
\ee
implies
\be
\eta_{1^r}(A)=r \sigma_{{_\square}}^{r-1} (A)\eta_{{_\square}}(A)
\ee

Another piece of information comes from \cite{MMS}.
Namely, in the HOMFLY case, when   $(\hbar=\overline{\hbar})$ we have
\be
\eta_R(A)+\overline{\eta}_R(A) =  \varkappa_R\sigma_{\square}^{|R|-2}\sigma_2(A)
\ee
where $\varkappa_R=\nu_{\overline R}-\nu_R$ and $\nu_R=\sum\limits_i r_i (i-1)$.
Here $r_i$ is a height of the column number $i$ in the Young diagram of the representation $R$.
Finally, $\sigma_2(A)$ is the second special polynomial, like $\sigma_{{_\square}}(A)$ it depends on the knot.

It is instructive to see how the reflection symmetry acts on the $\eta$-functions.
According to \cite{DMMSS} (see also \cite{GS} and references therein), it interchanges $R\leftrightarrow {\overline R}$
and also $q\leftrightarrow \frac{1}{t}$.
Then $\hbar \leftrightarrow -{\overline \hbar}$  and
the symmetry implies that
\be
\eta_R \rightarrow - {\overline \eta_{\overline R}}
\nonumber \\
{\overline \eta_R} \rightarrow - \eta_{\overline R}
\ee
Since for the fundamental representation ${\square} = {\overline {\square}}$,
we see that $\overline{ \eta_{\square}} = -\eta_{{\square}}$ and
\be
P_{\square}=\sigma_{\square} + \hbar \eta_{\square} + {\overline \hbar} {\overline \eta_{\square}} +\ldots
= \sigma_{\square} + \hbar \eta_{\square} - {\overline \hbar} {\eta_{\square}}+\ldots =
\sigma_{\square} - \hbar {\overline \eta_{\square}} + {\overline \hbar} {\overline \eta_{\square}} +\ldots
\ee

\bigskip
Now we can summarize what we know about symmetric and antisymmetric representations.
From $(3)$ we have
\be
P_{[r]}= \sigma_{[r]}+ \hbar \eta_{[r]} + {\overline \hbar} {\overline \eta_{[r]}}+\ldots
\ee
while from $(8)$ we know that
\be
\eta_{[r]}= \varkappa_R\sigma_{\square}^{|R|-2}\sigma_2 - {\overline\eta_{[r]}}
\ee
Substituting one into another, we get:
\be
\boxed{P_{[r]}=\sigma_{\square}^r+\overline{\eta}_{[r]}(\overline{\hbar}-\hbar)+\hbar\sigma_{\square}^{r-2}\sigma_2\varkappa_{[r]} + \ldots
= P^r_{{_\square}}+\hbar\varkappa_{[r]}\sigma_{\square}^{r-2}\sigma_2 + \ldots}
\ee
Similarly, for the antisymmetrical case:
\be
\boxed{P_{[1^r]}=\sigma_{\square}^r+\eta_{[1^r]}(\hbar-\overline{\hbar})-\overline\hbar\varkappa_{[1^r]}\sigma_{\square}^{r-2}\sigma_2 + \ldots
= P^r_{{_\square}}-\overline{\hbar}\varkappa_{[1^r]}\sigma_{\square}^{r-2}\sigma_2+ \ldots}
\ee

\bigskip
Now, let us compare these two formulas. They differ: one contains $\hbar$ and another $\bar\hbar$,
but now we can observe that for symmetric and antisymmetric representations
$\varkappa_R$ is rather special: since $\nu_{[r]}=0$,
$\varkappa_{[r]} = -\varkappa_{[1^r]} = \nu_{[1^r] } = \nu_{\overline{[r]}}$
and both formulas can be rewritten in a unified form:
\be
\boxed{P_R = P_{{_\square}}^{|R|} + (\hbar\nu_{\overline{R}}-\overline{\hbar}\nu_R)\,\sigma_{\square}^{|R|-2}\sigma_2+ \ldots
}
\ee
This is a remarkable formula, because in this form it can be used for
{\it arbitrary} representation, not obligatory symmetric and antisymmetric.
This formula is our new conjecture for the first deviation of arbitrary
superpolynomial from the special one.
At the moment there is no way to test this formula, because nothing is
known yet about the superpolynomials beyond (anti)symmetric representations.
For the first attempt to make use of $(16)$ -- in the case of
the figure-eight knot and $R=[2,1]$,
and with a somewhat controversial result -- see \cite{AMMM21}.
Further advances in this direction are very desirable.

\section*{Acknowledgements}
I am indebted to Alexei Morozov for useful discussions.
My work is partly supported by the
Ministry of Education and Science of the Russian Federation under the contract
8498,
by the grants
NSh-3349.2012.2,
RFBR-12-01-00525.

\end{document}